\documentclass[12pt]{article}
\usepackage{graphicx}
\usepackage{cite}
\usepackage{color}
\begin{document}
\title{The $Y(4140)$, $X(4260)$, $\psi(2D)$, $\psi(4S)$,\\
and tentative $\psi(3D)$}
\author{
Eef~van~Beveren$^{\; 1}$ and George~Rupp$^{\; 2}$\\ [10pt]
$^{1}${\small\it Centro de F\'{\i}sica Computacional,
Departamento de F\'{\i}sica,}\\
{\small\it Universidade de Coimbra, P-3004-516 Coimbra, Portugal}\\
{\small\it eef@teor.fis.uc.pt}\\ [10pt]
$^{2}${\small\it Centro de F\'{\i}sica das Interac\c{c}\~{o}es Fundamentais,}\\
{\small\it Instituto Superior T\'{e}cnico, Universidade T\'{e}cnica de
Lisboa,}\\
{\small\it Edif\'{\i}cio Ci\^{e}ncia, P-1049-001 Lisboa, Portugal}\\
{\small\it george@ist.utl.pt}\\ [.3cm]
{\small PACS number(s): 14.40.Gx, 13.25.Hw, 14.65.Dw, 11.80.Gw}
}

\maketitle

\begin{abstract}
Data on $B^{+}\rightarrow J/\psi\,\phi\,K^{+}$ and the $Y(4140)$ enhancement
recently reported by the CDF Collaboration \cite{ARXIV09032229}
are analyzed. The threshold behavior, as well as traces of the $X(4260)$
enhancement, the known $c\bar{c}$ resonances $\psi(2D)$, $\psi(4S)$, and a
tentative $\psi(3D)$ state, as observed in the mass distribution, suggest that
the $J/\psi\,\phi$ system has quantum numbers $J^{PC}=1^{--}$. It is then
argued that the $Y(4140)$ enhancement does not represent any kind of resonance,
but instead is a natural consequence of the opening of the $J/\psi\,\phi$
channel.
\end{abstract}

In Ref.~\cite{ARXIV09032229}, the CDF Collaboration displayed
the invariant-mass spectrum of $J/\psi\,\phi$
for the world's largest sample (75 events) of
exclusive $B^{+}\rightarrow J/\psi\phi K^{+}$ decays,
produced in $\bar{p} p$ collisions at $\sqrt{s}=1.96$ TeV,
and collected by the CDF-II detector at the Tevatron (Fermilab).
The experimental analysis revealed evidence for a narrow structure near
the $J/\psi\,\phi$ threshold, with significance in excess of 3.8$\sigma$.
Assuming an $S$-wave relativistic Breit-Wigner approximation,
the mass and width of this structure were determined as
$4143.0\pm2.9(\mathrm{stat})\pm1.2(\mathrm{syst})$ MeV/c,
and $11.7^{+8.3}_{-5.0}(\mathrm{stat})\pm3.7(\mathrm{syst})$ MeV/c,
respectively.

The $J/\psi\,\phi$ enhancement, which was baptized $Y(4140)$ by the CDF
Collaboration, has been studied in a variety of theoretical models,
namely as a $D_{\! s}^{\ast}\bar{D}_{\! s}^{\ast}$ molecule
with $J^{PC}=0^{++}$ or $J^{PC}=2^{++}$,
\cite{ARXIV09032529,ARXIV09033107,ARXIV09035424,ARXIV09041782,
ARXIV09054178}
or an exotic hybrid charmonium state with $J^{PC}=1^{-+}$
\cite{ARXIV09033107,ARXIV09035200}. It was also shown that the
$Y(4140)$ is probably not the second radial excitation
of any of the $P$-wave charmonium states \cite{ARXIV09040136}.
Finally, in Ref.~\cite{ARXIV09035200} it was argued that the
$Y(4140)$ cannot be a $D_{\! s}^{\ast}\bar{D}_{\! s}^{\ast}$ molecule.
Surprisingly, no tetraquark proposals have circulated yet.

The CDF Collaboration observed that the $J/\psi\,\phi$ enhancement
is well above the threshold for open-charm decays,
and a $c\bar{c}$ charmonium resonance with this mass
would be expected to decay dominantly into an open-charm pair,
with only a tiny branching fraction into $J/\psi\,\phi$
\cite{PRD69p094019,RMP80p1161}.
Consequently, according to the CDF Collaboration,
this structure is not compatible with conventional expectations
for a charmonium resonance.

Indeed, the $Y(4140)$ enhancement lies relatively close to
open-charm thresholds, namely $D_{\! s}D_{\! s}^{\ast}$,
$D^{\ast}D^{\ast}$, $D_{\! s}D_{\! s}$, and $DD^{\ast}$, at
4.076, 4.02, 3.939, and 3.875~GeV, respectively.
However, branching fractions have been measured at 4.028~GeV,
with the SLAC/LBL magnetic detector at SPEAR \cite{PLB69p503}.
The results suggest that the opening of a channel
is followed by a rather fast falling-off
at higher invariant masses \cite{ARXIV09051595}.
For conservative parameters, we find that at 4.14~GeV
the $DD^{\ast}$ channel has almost completely faded out,
whereas the other three open--charm channels, viz.\
$D_{\! s}D_{\! s}$, $D^{\ast}D^{\ast}$, and $D_{\! s}D_{\! s}^{\ast}$,
are reduced to less than 2\%, 10\%, and 25\%, respectively.
Taking into account the combinatorial factors as well,
we find that actually not so much open-charm decay should be expected
at 4.14~GeV.

A related observation,
for pion multiplicities in proton-antiproton annihilation
\cite{PTP72p1050,ZPC67p281},
shows that a $p\bar{p}$ system at rest does not preferably decay into
pairs of pions, but instead into several more pions, with a maximum of five,
probably through other resonances \cite{NPB30p525}.
Apparently, excess of kinetic energy does not contribute
to the conditions in which resonances can be formed.
Here, we study the formation of a $J/\psi\,\phi$ pair of resonances,
which we thus expect to be at maximum just above threshold.
Similar phenomena have been observed for
$J/\psi\,\rho$ \cite{PRL93p072001},
$J/\psi\,f_{0}(980)$ \cite{PRL95p142001,ARXIV08081543},
$\psi (2S)f_{0}(980)$ \cite{PRL99p142002},
and $\Lambda_{c}\bar{\Lambda}_{c}$ \cite{PRL101p172001}.

\begin{figure}[htbp]
\begin{center}
\begin{tabular}{c}
\scalebox{0.8}{\includegraphics{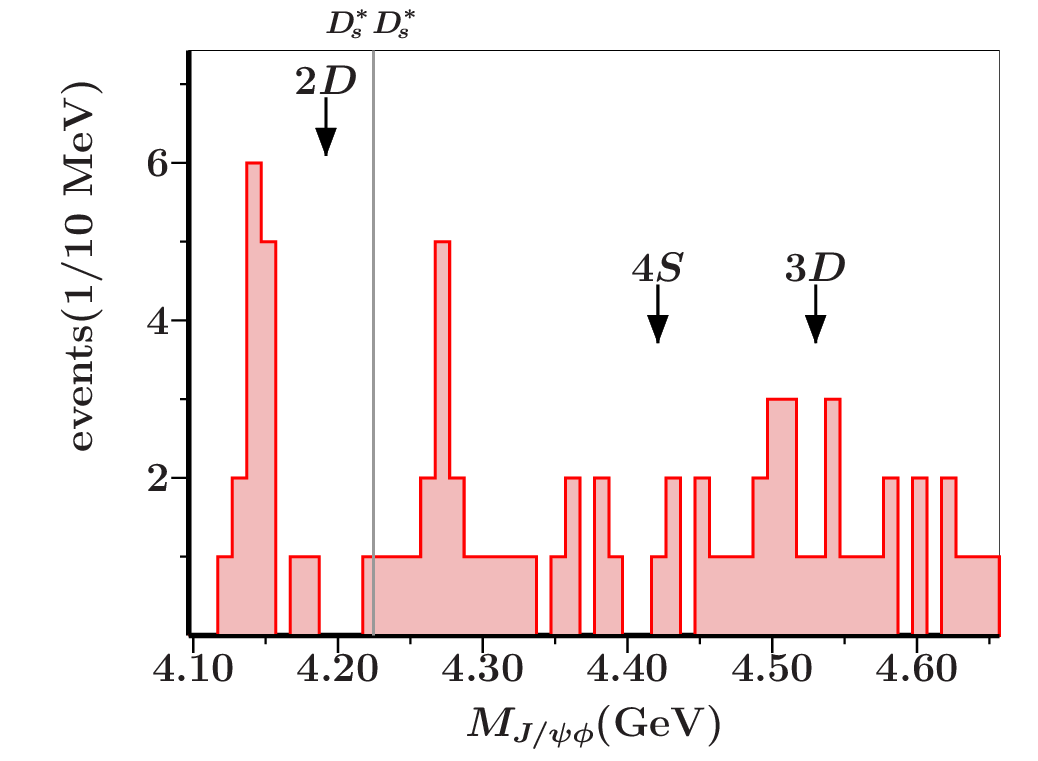}}\\ [-15pt]
\end{tabular}
\end{center}
\caption[]{\small
Experimental invariant-mass distribution of $J/\psi\,\phi$
for $B^{+}$ decays into $J/\psi\phi K^{+}$,
from the CDF Collaboration \cite{ARXIV09032229}.
With arrows we indicate the central-mass positions
of the well-established
$\psi (2D,4160)$ and $\psi (4S,4415)$
$c\bar{c}$ resonances \cite{PLB667p1,PLB660p315},
as well as the recently discovered
$\psi(3D,$4530--4550) \cite{EPL85p61002,ARXIV09044351}
$c\bar{c}$ state.
Moreover, the threshold energy \cite{PLB667p1} of the
$D_{\! s}^{\ast}\bar{D}_{\! s}^{\ast}$ channel
is shown.
}
\label{cdf}
\end{figure}

In Fig.~\ref{cdf}, we display the events recorded by the CDF Collaboration
\cite{ARXIV09032229}.
The signal shows a clear and narrow peak, which sets out
at the $J/\psi\phi$ threshold.
Some minor structures at about 4.27,
4.40, and 4.53~GeV are less significant.

The weak decay $B^{+}\rightarrow Y(4140)\,K^{+}$
does not necessarily conserve parity,
so the final $Y(4140)\,K^{+}$ state may have
either $J^{P}=0^{-}$ or $J^{P}=0^{+}$.
In the $J^{P}=0^-$ scenario, since the $K^{+}$ is also a pseudoscalar, the
simplest assignments for the $Y(4140)$ are a scalar in a relative $S$
wave with the $K^{+}$, or a vector in a $P$ wave. On the other hand, if
$Y(4140)\,K^{+}$ has $J^{P}=0^{+}$, then the $Y(4140)$ can be either a
pseudoscalar in a relative $S$ wave with the $K^{+}$, or an axial vector in
a $P$ wave. Also, the $Y(4140)$ decays into a pair of vector particles.
For a scalar or axial vector, this can be either in an $S$ wave or a
$D$ wave, whereas for a pseudoscalar or vector it can only be in a $P$ wave.
Moreover, from the data of the CDF Collaboration \cite{ARXIV09032229}, and
assuming a natural opening of the $J/\psi\,\phi$ channel, as well as
the absence of resonances in the $J/\psi\,\phi$ system in the relevant energy
interval, we conclude from the behavior of the signal near the $J/\psi\phi$
threshold that the system is most probably in a relative
$P$ wave \cite{MorseFeshbach}.  If indeed so, we are left with the assignments
$J^{P}=0^{-}$ and $J^{P}=1^{-}$ for the $Y(4140)$. Furthermore, there seem to
be significant structures in the CDF \cite{ARXIV09032229} data, displayed in
Fig.~\ref{cdf}, near the $\psi(4S,4415)$ and tentative
\mbox{$\psi(3D,$4530--4550)} \cite{EPL85p61002,ARXIV09044351} vector
resonances, thus suggesting $J^{P}=1^{-}$ for the $J/\psi\,\phi$ system.

We are well aware that the CDF Collaboration assumed an $S$ wave for the
$J/\psi\,\phi$ system in Ref.~\cite{ARXIV09032229}. Nevertheless, based on our
observations above, we are more inclined to choose vector quantum numbers as
our working hypothesis. Moreover, there appears to be an enhancement at about
4.27~GeV in Fig.~\ref{cdf}.
This coincides with the invariant-mass region where,
by studying the $e^{+}e^{-}\to J/\psi\,\pi^{+}\pi^{-}$ cross section,
the BABAR Collaboration discovered a new vector enhancement,
originallly baptized as $Y(4260)$ \cite{PRL95p142001},
but now included in the PDG tables as $X(4260)$ \cite{PLB667p1}.
If the structure observed here indeed corresponds to the $X(4260)$, then our
vector assignment for the $J/\psi\,\phi$ system is quite reasonable.

Next we shall discuss whether the $Y(4140)$ enhancement should be interpreted
as a new resonance. Resonances are characterized by complex poles in the
scattering amplitude \cite{PRD21p772,PRD27p1527,ZPC30p615}.
From Refs.~\cite{PRD21p772,PRD27p1527} we learn that the $c\bar{c}$ pole
spectrum is very rich, and has in the mass region 4.0--4.2~GeV some 11
resonances, for $J^{PC}=(0,2,4)^{-+}$ and $J^{PC}=(1,2,3,4,5)^{--}$,
where, moreover, $J^{PC}=(1,3,5)^{--}$ can have two different angular
excitations. Furthermore, {\it dynamically generated} \/poles may show up
as well \cite{PRD76p074016}, just like the dynamically generated nonet of
light scalar mesons \cite{ZPC30p615,AIPCP1030p219},
and the variety of examples for open charm and beauty
\cite{HEPPH0312078,MPLA19p1949,PRL91p012003}.

Unfortunately, most of the $c\bar{c}$ resonances are still awaiting discovery
in experiment.  However, no matter if observed or not, we expect a very rich
charmonium spectrum for higher invariant masses. Adding to that possible
tetraquarks, molecules, and hybrids would saddle us up with an approximate
continuum of resonances. However, do not support such a classification
scheme of cross-section peaks showing up in experiment.
We tend to believe that not every bump represents a new resonance, besides the
fact that true resonances may have very different appearances in experiment,
and even may manifest themselves through the absence of any signal
\cite{ARXIV08111755,ARXIV09044351}. In particular, if we take for the resonance
position of the $\psi (2D,4160)$ the latest result of the BES Collaboration,
namely $4191.7\pm 6.5$ MeV \cite{PLB660p315},
we see that the average mass of the $Y(4140$ strucuture
lies well in between the peaks
of the $\psi (3S,4040)$ and $\psi (2D,4160)$ resonances.
As for open-charm decay, we expect its main influence
near the central resonance masses,
where the $c\bar{c}$ system resonates most strongly.
We shall come back to this point later on.

In Ref.~\cite{ARXIV09051595}, we discussed two distinct modes
for the decay of a $c\bar{c}$ system. One is through string breaking, giving
rise to OZI-allowed pair creation, in which the $c$ quark and $\bar{c}$
antiquark recombine with the created quark and antiquark, resulting in
charm-meson decay.  The other mode corresponds to --- OZI-forbidden ---
$q\bar{q}$ emission, probably taking place in the gluon cloud surrounding the
charmed constituents. In such processes, the $c\bar{c}$ propagator radiates off
systems of light quarks, and then jumps to its ground state \cite{PRL95p142001}
or to a lower-lying excitation \cite{PRL99p142002}.
It is well known that OZI-forbidden processes are suppressed with respect
to OZI-allowed ones \cite{OZI}. In Ref.~\cite{ARXIV09051595}, we tried
to quantify the difference in probability for these two distinct processes.
However, it is important to notice that coupling to virtual $q\bar{q}$ pairs in
the gluonic periphery may happen at any energy, so that no specific mass can be
associated with it.

In the present case, we are considering the formation of pairs of $J/\psi$ and
$\phi$ resonances. According to the prior discussion, this is most likely to
happen just above threshold, via an $s\bar{s}$ quark-antiquark pair created in
the peripheral glue. Such an $s\bar{s}$ pair couples to a dressed $s\bar{s}$
propagator that contains all $s\bar{s}$ resonances, viz.\
$\eta$, $\eta '$, $f_{0}(980)$, $\phi$, and so forth \cite{ARXIV08091149}.
By concentrating on $K^{+}K^{-}$ in the experimental analysis
\cite{ARXIV09032229}, one will thus  first observe
the lowest possible resonance in this channel, which is the $\phi$.
There actually is little surprise in such a result.
But its study is very important, and may certainly help in leading us
to a theory for the formation of hadrons.  Let us next focus on the results
of the CDF Collaboration, displayed in Fig.~\ref{cdf}.

\begin{figure}[htbp]
\begin{center}
\begin{tabular}{c}
\scalebox{0.8}{\includegraphics{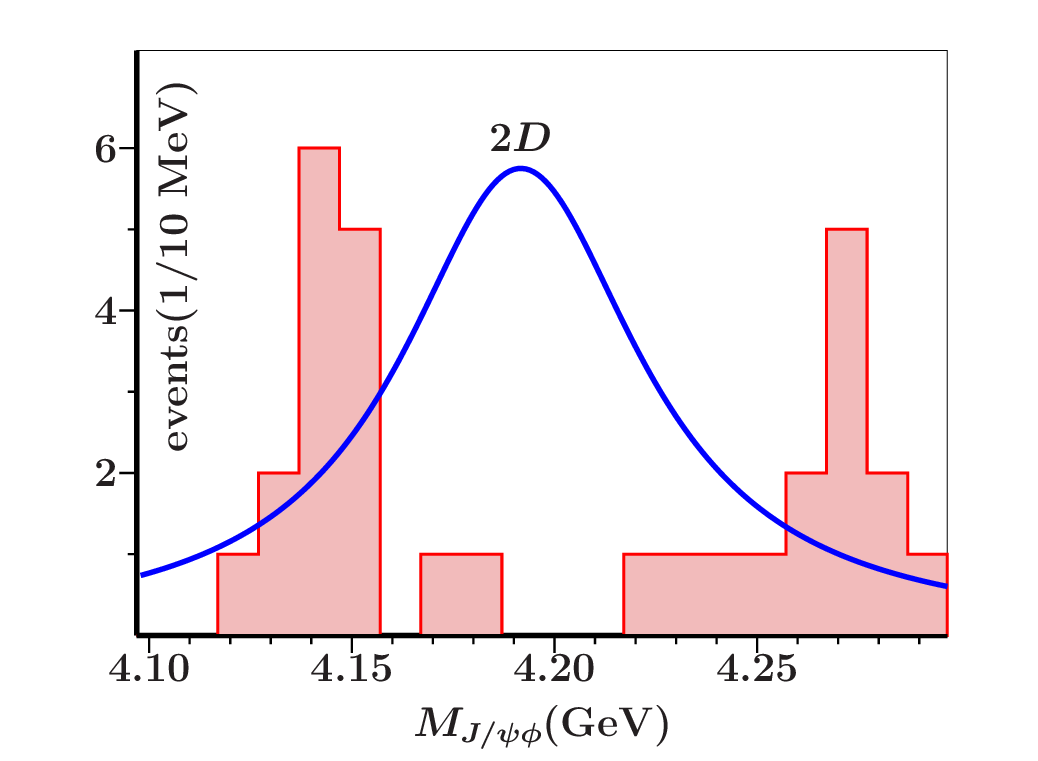}}\\ [-15pt]
\end{tabular}
\end{center}
\caption[]{\small
Experimental invariant-mass distribution of $J/\psi\,\phi$,
for $B^{+}$ decays into $J/\psi\phi K^{+}$,
from the CDF Collaboration \cite{ARXIV09032229}.
The solid line (blue) stands for a simpel Breit-Wigner approximation
of the $\psi (2D,4160)$ \cite{PLB660p315} resonance.
}
\label{cdf2}
\end{figure}

The signal sets out from the $J/\psi\,\phi$ threshold at
$4116.37\pm 0.03$~GeV \cite{PLB667p1} upwards,
and seems to rise linearly towards its maximum.
This indicates a relative $P$ wave, as we already stated in the foregoing,
so $S$ and $D$ waves seem to be excluded. This implies that $J=0$ or $J=2$,
assumed for molecular explanations
\cite{ARXIV09032529,ARXIV09033107,ARXIV09035424,ARXIV09041782,ARXIV09054178},
are not very plausible. However, the signal never reaches its true maximum,
because of the presence of the $\psi(2D,4160)$ resonance.
In Fig.~\ref{cdf2} we have depicted the situation.

From Fig.~\ref{cdf2} it is clear that some $J/\psi\,\phi$ signal is missing
exactly where one expects many charm-meson pairs to be formed, viz.\ around
the peak of the $\psi (2D,4160)$ resonance. In
Refs.~\cite{ARXIV08111755,ARXIV09044351},
we argued that the latter process takes place
via string breaking, and is moreover faster than
peripheral emission of $s\bar{s}$ pairs.
Consequently, open-charm decays deplete the $c\bar{c}$ propagator well
before it can decay into $J/\psi\,\phi$.
However, another important observation can be made here. Namely, there
is quite some spreading \cite{PLB667p1} in the experimental results for
the mass and width of the $\psi (2D,4160)$. Here, we have used the values
of Ref.~\cite{PLB660p315}, which seem to agree well with the data of the
CDF Collaboration.  Hence, a precision experiment for the $J/\psi\,\phi$
signal could be an adequate alternative method to establish
the mass and width of this charmonium resonance.

For the feeble signal in the 4.26~GeV region,
exactly the same arguments hold as for the $X(4260)$,
which we presented in Ref.~\cite{ARXIV09051595}.
Namely, near the threshold of the OZI-allowed
$D_{\! s}^{\ast}\bar{D}_{\! s}^{\ast}$ channel, the $c\bar{c}$ propagator
is in an $s\bar{s}$-rich environment. The latter also couples to the
$s\bar{s}$ propagator, probably most to the $f_{0}(980)$ part,
but certainly also to the $\phi$. On the one hand, $s\bar{s}$ creation in
the inner region near the fast oscillating $c\bar{c}$ pair, takes signal
away from the peripheral process $c\bar{c}\to J/\psi\,\phi$, via open-charm
decay. But on the other hand, enhanced $s\bar{s}$ creation in the inner
region due to opening of the $D_{\! s}^{\ast}\bar{D}_{\! s}^{\ast}$ channel
gives rise to a higher probability of $s\bar{s}$ pairs to escape from that
region. For the $X(4260)$ in $J/\psi\,\pi\pi$, this leads to a positive
net balance \cite{ARXIV09051595}, and so probably also for $J/\psi\,\phi$.

As far as the remaining part of the $J/\psi\,\phi$ mass distribution
is concerned, we may observe from Fig.~\ref{cdf} that some effect of the
negative interference of the $\psi(4S,4415)$ is visible, albeit with little
statics. Nevertheless, at the place where we predicted
\cite{EPL85p61002,ARXIV09044351} the $\psi(3D,$4530--4550) resonance, some
signal is clearly visible, which seems to confirm our prediction.

In summary, we may conclude that the $J/\psi\,\phi$ system
in exclusive $B^{+}\rightarrow J/\psi\phi K^{+}$ decays
probably has quantum numbers $J^{PC}=1^{--}$, and that the bump at 4.14~GeV
does not represent any resonance. The results of the CDF Collaboration
are nonetheless a very valuable contribution to our understanding
of strong interactions and hadron formation.

We have also indicated how and where previously observed structures,
viz.\ the $X(4260)$, $\psi(2D)$, $\psi(4S)$, and the preliminary $\psi(3D)$,
show up in the data of the CDF Collaboration.

We are grateful for the rather precise measurements
of the CDF Collaboration, which made the present analysis possible.
This work was supported in part by
the \emph{Funda\c{c}\~{a}o para a Ci\^{e}ncia e a Tecnologia}
\/of the \emph{Minist\'{e}rio da Ci\^{e}ncia, Tecnologia e Ensino Superior}
\/of Portugal, under contract no.\ CERN/\-FP/\-83502/\-2008.

\newcommand{\pubprt}[4]{{#1 {\bf #2}, #3 (#4)}}
\newcommand{\ertbid}[4]{[Erratum-ibid.~{#1 {\bf #2}, #3 (#4)}]}
\def\AIPCP{AIP Conf.\ Proc.}
\def\EPL{Europhys.\ Lett.}
\def\MPLA{Mod.\ Phys.\ Lett.\ A}
\def\NPB{Nucl.\ Phys.\ B}
\def\PLB{Phys.\ Lett.\ B}
\def\PRD{Phys.\ Rev.\ D}
\def\PRL{Phys.\ Rev.\ Lett.}
\def\PTP{Prog.\ Theor.\ Phys.}
\def\RMP{Rev.\ Mod.\ Phys.}
\def\ZPC{Z.\ Phys.\ C}

\end{document}